\documentclass[12pt]{article}

\usepackage{amssymb}
\usepackage{amsmath}
\usepackage{graphicx,color}
\usepackage[nosort]{cite}

\def\dfrac#1#2{\displaystyle\frac{#1}{#2}}

\newcommand{\arxiv}[1]{arXiv:#1}
\newcommand{\tr}{{\rm tr}}
\newcommand{\PpT}{(P^+_\mu)^\mathrm{T}}
\newcommand{\PmT}{(P^-_\mu)^\mathrm{T}}
\newcommand{\U}{\mathrm{U}}
\newcommand{\T}{\mathrm{T}}
\newcommand{\mrm}{\mathrm}
\newcommand{\iu}{\mathrm{i}}
\newcommand{\e}{{\hspace{.1em}{\mathrm e}}}

\newcommand{\A}{X}
\newcommand{\B}{Y}
\allowdisplaybreaks

\begin{document}

\baselineskip=17pt

\begin{titlepage}
\rightline{\tt RIKEN-MP-36}
\rightline{\tt UTHEP-635}
\rightline{\tt UT-Komaba/11-11 }
\rightline{\tt YITP-11-91}
\begin{center}
\vskip 0.5cm
{\Large \bf {Revisiting symmetries of lattice fermions
}}\\
\vskip 0.4cm
{\Large \bf {via spin-flavor representation}}
\vskip 0.9cm
{ {Taro Kimura$^{a,b\,}$\footnote{kimura@dice.c.u-tokyo.ac.jp},
Shota Komatsu$^{c\,}$\footnote{skomatsu@hep1.c.u-tokyo.ac.jp},
Tatsuhiro Misumi$^{d\,}$\footnote{misumi@yukawa.kyoto-u.ac.jp},}}
\vskip 0.5cm
{ {Toshifumi Noumi$^{c\,}$\footnote{tnoumi@hep1.c.u-tokyo.ac.jp},
Shingo Torii$^{c\,}$\footnote{storii@hep1.c.u-tokyo.ac.jp},
Sinya Aoki$^{e\,}$\footnote{saoki@het.ph.tsukuba.ac.jp}}}
\vskip 0.7cm
{\small
{\it {$^a$ Department of Basic Science, University of Tokyo,
Tokyo 153-8902, Japan}}\\
{\it {$^b$ Mathematical Physics Lab., RIKEN Nishina Center, Saitama 351-0198, Japan}}\\
{\it {$^c$ Institute of Physics, University of Tokyo,
Tokyo 153-8902, Japan}}\\
{\it {$^d$Yukawa Institute for Theoretical Physics, Kyoto University,
Kyoto 606-8502, Japan}}\\
{\it {$^e$Graduate School of Pure and Applied Sciences, 
University of Tsukuba, Ibaraki 305-8571, Japan}}}\\[17mm]

{\bf Abstract}
\end{center}

\noindent
Employing the spin-flavor  representation, 
we investigate the structures of the doubler-mixing symmetries and 
the mechanisms of their spontaneous breakdown in four types 
of lattice fermion formulation.
We first revisit the $\U(4)\times\U(4)$ symmetries of the naive fermion 
with the vanishing bare mass $m$, and re-express
them in terms of the spin-flavor representation.
We apply the same method to the Wilson fermion, 
which possesses only the $\U(1)$ vector symmetry for general 
values of $m$. For a special value of $m$, however, there emerges an 
additional $\U(1)$ symmetry to be broken by pion condensation. 
We also explore two types of minimally doubled fermion, and 
discover a similar kind of symmetry enhancement and its 
spontaneous breakdown.

\end{titlepage}

\newpage

%%%%%%%%%%   Introduction   %%%%%%%%%%

\section{Introduction}
\label{sec:Intro}
\setcounter{footnote}{0}

Lattice field theory is the most powerful theoretical framework for 
investigating non-perturbative aspects of gauge theories such as 
quantum chromodynamics (QCD). In this framework, while gauge fields 
are successfully incorporated without any obstacles \cite{Wil}, putting fermion fields on 
the lattice is not straight-forward: A naive fermion formulation 
gives rise to undesirable superfluous fermion modes, producing 
sixteen flavors rather than one. The essence of this problem is summarized in the 
form of Nielsen-Ninomiya's no-go theorem \cite{Karsten:1980wd, NN}, which states that a 
lattice fermion action with chiral symmetry, locality and other 
reasonable properties inevitably produces   one 
or more pairs of particles with opposite chiralities in the continuum limit.

To overcome this doubling problem, various types of lattice fermion 
formulation have been proposed. Among them, the staggered fermion 
\cite{KS,Suss,Sha} and the Wilson fermion \cite{Wil} are utilized frequently 
for QCD simulations. In the staggered fermion formulation, the number 
of flavors is reduced to four by the spin diagonalization method 
\cite{KawaS}.
The staggered fermion action possesses the flavor-singlet vector and non-singlet axial-vector $\U(1)$ symmetries, the latter of which is spontaneously broken. In addition, it has several discrete 
symmetries as well \cite{GS}.  
In the case of the Wilson fermion, on the other hand, doublers are eliminated by  adding the Wilson term at the cost of the chiral symmetry:
The term is not invariant under the chiral transformation even if the bare quark mass is set to zero,
so that the additive mass renormalization is required to realize massless or very light quarks. 
At the same time, the Wilson term splits 
sixteen species doublers of the naive fermion into five branches 
composed of one, four, six, four, and one fermion modes.
One of the notable features of the Wilson 
fermion is its complicated phase structure \cite{AokiP, AokiU1, CreutzW, SS}:
Parity symmetry is spontaneously broken by pion condensation, 
depending on the values of the quark mass and the gauge coupling constant.

As is seen in the above,  the existence of  fermion doublers on the lattice entails the symmetry 
structure different from that of continuum theory. 
Hence, it is important  to understand the underlying symmetries in lattice field theory,
not only  for QCD simulations with these fermion formulations,
but also  for constructions of other lattice fermion formulations such as overlap \cite{GW,Neu}, domain-wall \cite{Kap,Shamir,FuSh}, 
staggered Wilson \cite{CKM2, Adams1, Adams2, Hoel}, staggered overlap \cite{Adams1, Adams2, CKM1, PdF}  and minimally doubled fermions \cite{KW, CB, BBTW, KM,CCW, CCWW, Cichy, CMK, CreutzPS}.  
Moreover fermion doublers could be used to economically simulate QCD-like theories with many flavors,\footnote{ See, for example, 
refs.~\cite{Appel, Deuz, Onogi, Hasen, Hayakawa, DeGrand, Debbio}} which might be relevant for the construction of techni-color theories.

In the case of the naive fermion, 
it is known that the kinetic term possesses 
$\U(4)\times \U(4)$ symmetries\footnote{
 For earlier discussion on the symmetry of the naive fermion, see also \cite{Drell, Blairon}.} and it is spontaneously broken into the diagonal $\U(4)$~\cite{KawaS, KMN1}.
In the case of the Wilson fermion,  branches other than the physical one have been taken little heed of. 
Recently, it has been reported in the study of the 
Gross-Neveu model that the symmetry enhancement would take place at the 
central branch (the third branch) of Wilson-type fermions \cite{CKM2}.

The aim of this paper is to shed light on the structures of underlying continuous 
symmetries and their spontaneous breakdown in four types of lattice fermions 
formulation: the naive fermion, the Wilson fermion and two kinds of minimally 
doubled fermion. For this purpose, we rewrite lattice fermion actions in 
``the spin-flavor representation'' \cite{KMN3, Gliozzi}, in which 
the spin and doubler-multiplet 
structures of the lattice fermions become manifest.  
We first
re-express the $\U(4)\times \U(4)$ symmetry of the naive fermion in~\cite{ KawaS, KMN1}
using the spin-flavor representation. 
We then apply the same method to the Wilson fermion action,
which is invariant under only the ordinary $\U(1)$ vector transformation for general 
values of the mass parameter $m$. 
We show, however, that an additional $\U(1)$ vector symmetry is realized by tuning $m$ and
this symmetry is spontaneously broken by pion condensation. 
Finally, we 
explore the Karsten-Wilczek and the Bori\c{c}i-Creutz minimally doubled fermion
and discover that a similar type of symmetry enhancement and its spontaneous 
breakdown occur. 

This paper is organized as follows. In section \ref{sec:NF}, we revisit the symmetries 
of the naive lattice fermion via the spin-flavor representation.
In section \ref{sec:WF}, we discuss the symmetries of the Wilson fermion 
with emphasis on the symmetry enhancement and its spontaneous breakdown. 
We also explore minimally doubled fermions in section \ref{sec:MDF}. Section \ref{sec:SD} is devoted to a summary and discussions. Some technical details are given in appendixes.

%%%%%%%%%%%%%% Naive fermion %%%%%%%%%%

\section{Naive fermion and Spin-flavor representation}
\label{sec:NF}
In this section, we first review the U$(4)\times$U$(4)$ symmetries of the naive fermion \cite{KawaS,KMN1}. Then we introduce the spin-flavor representation, which simplifies the identification of symmetry in the case of the Wilson fermion and the minimally doubled fermions.  

The action of the naive fermion is given by
\begin{equation}
\label{naive_action}
S_{\mrm{nf}}=\frac{1}{2} \sum _{n,\mu} (\bar{\psi}_{n}\gamma _{\mu}\psi _{n+\hat{\mu}
} -\bar{\psi}_{n+\hat{\mu}}\gamma _{\mu}\psi _{n} )+m\sum_{n}\bar{\psi}_n\psi_n\,.
\end{equation}
Throughout this paper, we consider the nondimensionalized action. As is discussed in \cite{KawaS,KMN1}, the kinetic term of this action has larger symmetry than the action of the continuum theory:
\begin{equation}
\label{sym_naive_all}
\begin{split}
&\psi_n\rightarrow\psi_n^\prime=\exp\Big[\iu \sum _\A \left(\theta _\A^{(+)}\Gamma^{(+)}_\A+\theta _\A^{(-)}\Gamma^{(-)}_\A\right)\Big]\psi_n\,,\\
&\bar{\psi}_n\rightarrow\bar{\psi}_n^\prime=\bar{\psi}_n\exp\Big[\iu\sum _\A\left(-\theta _\A^{(+)}\Gamma^{(+)}_\A+\theta _\A^{(-)}\Gamma^{(-)}_\A\right)\Big]\,.
\end{split}
\end{equation}
Here, $\Gamma^{(+)}_\A$ and $\Gamma^{(-)}_\A$ are site-dependent $4\times 4$ matrices:
\begin{eqnarray}
\label{sym_t+}
\Gamma^{(+)}_\A&\in &\left\{\mathbf{1}_4\,,\,\, (-1)^{n_1+\ldots+ n_4}\gamma_5\,,\,\,(-1)^{\check{n}_\mu}\gamma_\mu\,,\,\,(-1)^{n_\mu}\iu \gamma_\mu\gamma_5
\,,\,\,(-1)^{n_{\mu,\nu}}\frac{\iu \,[\gamma_\mu\,,\gamma_\nu]}{2}\right\}\,,\\
\label{sym_t-}
\Gamma^{(-)}_\A&\in &\left\{(-1)^{n_1+\ldots+ n_4}\mathbf{1}_4\,,\,\, \gamma_5\,,\,\,(-1)^{n_\mu}\gamma_\mu\,,\,\,(-1)^{\check{n}_\mu}\iu \gamma_\mu\gamma_5
\,,\,\,(-1)^{\check{n}_{\mu,\nu}}\frac{\iu \,[\gamma_\mu\,,\gamma_\nu]}{2}\right\}\,,
\end{eqnarray}
where $\check{n}_\mu=\sum _{\rho\neq\mu}n_{\rho}$, $n_{\mu,\nu}=n_\mu+n_\nu$ and $\check{n}_{\mu,\nu}=\sum_{\rho\neq\mu,\nu}n_\rho$.
Although the kinetic term is invariant under the transformations with arbitrary complex $\theta_\A^{(\pm)}\,$s, the link reflection positivity constrains $\theta_\A^{(\pm)}\,$s to be real. In other words, only if $\theta_\A^{(\pm)}\,$s are real numbers, the transformations commute with the following anti-linear operation $\Theta$:  
\begin{equation}
\Theta[\psi_n]=\bar{\psi}_{n_i,-n_4+1}\,\gamma_4\,,\quad
\Theta[\bar{\psi}_n]=\gamma_4\psi_{n_i,-n_4+1}\,.
\end{equation}

The symmetry group (\ref{sym_naive_all}) is $\U(4)\times \U(4)$\footnote{The so-called ``doubling symmetry" \cite{Karsten:1980wd} is a discrete subgroup of $\U(4)\times \U(4)$.}, which is broken by the chiral condensate or the mass term down to the diagonal $\U(4)$ generated by $\Gamma_\A^{(+)}$. Therefore, there appear sixteen Nambu-Goldstone bosons (NG bosons) when the symmetry is spontaneously broken. The existence of these sixteen NG bosons is explicitly verified from the strong coupling analysis~\cite{KMN1,KMN2}. 

In order to understand
the relation between the $\U(4)\times \U(4)$ symmetries
and the sixteen doublers,
it is useful to rephrase the above results
using the spin-flavor representation \cite{KMN3,Gliozzi}.
We use this representation also in the subsequent sections
to clarify the symmetries of the Wilson and minimally doubled fermions.

Using the field $\chi_n$ defined by
\begin{equation}
\chi_n=\gamma _4^{n_4} \gamma _3 ^{n_3}\gamma _2^{n_2} \gamma _1^{n_1}\psi_n,
\quad
\bar{\chi}_n=\bar{\psi}_n\gamma _1^{n_1} \gamma _2 ^{n_2}\gamma _3^{n_3} \gamma _4^{n_4}\,,
\end{equation}
we can represent the naive fermion action (\ref{naive_action}) as follows:
\begin{equation}
\label{spin-diagonalized_action}
S_{\mrm{nf}}=\left[\sum_{n,\mu}\eta_\mu(n) \,\bar{\chi}_n\,\frac{\chi_{n+\hat{\mu}}-\chi_{n-\hat{\mu}}}{2}+m\sum_{n}\bar{\chi}_n\chi_n\right]\quad {\rm with}\quad \eta_\mu(n)=(-1)^{\sum_{\nu<\mu}n_\nu}\,.
\end{equation}
As in the case of the staggered fermion \cite{KMN3,Gliozzi}, we define the field $\Psi(N)$ as\footnote{
The spin-flavor representation becomes more complicated
in the presence of gauge fields.
However,
because the structure of global symmetry remains unchanged,
we here consider the free theory for simplicity.
}
\begin{equation}
\Psi(N)_{\alpha,\,f_1,\,f_2}=\sum_{A}\left(\frac{\gamma_A}{2}\right)_{\alpha,\,f_1}\chi_A(N)_{f_2}\,,
\quad
\bar{\Psi}(N)_{\alpha,\,f_1,\,f_2}=\sum_{A}\left(\frac{\bar{\gamma}_A}{2}\right)_{\alpha,\,f_1}\bar{\chi}_A(N)_{f_2}\,,
\end{equation}
where $A_\mu=0$ or 1, $\chi_A(N)=\chi_{2N+A}$, $\gamma_A=\gamma_1^{A_1}\gamma_2^{A_2}\gamma_3^{A_3} \gamma_4^{A_4}$, and $\bar{\gamma}_A$ denotes the complex conjugate of $\gamma_A$.
Note that we can identify the index $\alpha$ with a spinor index and the indices $f_1$, $f_2$ with flavor indices of sixteen doublers.
The relation between the fields $\psi$ and $\Psi$ is given by
\begin{eqnarray}
\label{sfd1}
\Psi(N)_{\alpha,\,f_1,\,f_2}=\sum_{A}\left(\frac{\gamma_A}{2}\right)_{\alpha,\,f_1}
\left(\bar{\gamma}_A\right)_{\beta,\,f_2}\left(\psi_{2N+A}\right)_\beta\,,\label{Psi}\\
\label{sfd2}
\bar{\Psi}(N)_{\alpha,\,f_1,\,f_2}=\sum_{A}\left(\frac{\bar{\gamma}_A}{2}\right)_{\alpha,\,f_1}
\left(\gamma_A\right)_{\beta,\,f_2}\left(\bar{\psi}_{2N+A}\right)_\beta\,.\label{Psibar}
\end{eqnarray}
In terms of $\Psi(N)$,
the naive fermion action (\ref{naive_action})
can be written as
\begin{eqnarray}
\nonumber
S_{\mrm{nf}}&=&\Bigg[\frac{1}{2}\sum_{N,\mu}\bar{\Psi}(N)\left(\gamma_\mu\otimes\mathbf{1}_4\otimes\mathbf{1}_4\right)\nabla_\mu\Psi(N)
+m\sum_{N}\bar{\Psi}(N)\left(\mathbf{1}_4\otimes\mathbf{1}_4\otimes\mathbf{1}_4\right)\Psi(N)\\
\label{naive_sfd}
&&\hspace{8mm}
+\frac{1}{2}\sum_{N,\mu}\bar{\Psi}(N)\left(\gamma_5\otimes\gamma_\mu^\T\gamma_5^\T\otimes\mathbf{1}_4\right)\nabla^2_\mu\Psi(N)\Bigg]\,,
\end{eqnarray}
where
\begin{eqnarray}
\bar{\Psi}(A\otimes B \otimes C)\Psi&=&\sum_{\alpha,\,\alpha^\prime,\,f_1,\,f_1^\prime,\,f_2,\,f_2^\prime}
\bar{\Psi}_{\alpha,f_1,f_2}\,(A)_{\alpha\alpha^\prime}\,(B)_{f_1f_1^\prime}\,(C)_{f_2f_2^\prime}\,\Psi_{\alpha^\prime,f_1^\prime,f_2^\prime}\,,\\
\nabla_\mu \Psi(N)&=&\frac{\Psi(N+\hat{\mu})-\Psi(N-\hat{\mu})}{2}\,,\label{nabla}\\
\nabla^2_\mu \Psi(N)&=&\frac{\Psi(N+\hat{\mu})-2\Psi(N)+\Psi(N-\hat{\mu})}{2}\,,\label{nabla2}
\end{eqnarray}
and the superscript $\T$ denotes transposition.
In (\ref{naive_sfd}), the third term 
breaks the vector and the axial-vector symmetries among sixteen doublers
down to the $\U(4)\times \U(4)$ symmetries
generated by $(\mathbf{1}_4\otimes \mathbf{1}_4\otimes \mathfrak{u}(4))$
and $(\gamma_5\otimes \gamma_5^T\otimes \mathfrak{u}(4))$.
Thus, the $\U(4)\times\U(4)$ symmetries are part of the vector and the axial-vector symmetries among doublers. 
The reformulated action (\ref{naive_sfd}) is quite similar to that of the staggered fermion \cite{KMN3,Gliozzi} except that eq.~(\ref{naive_sfd}) has an additional tensor structure denoted by the index $f_2$. This is manifestation of the well-known fact that the naive fermion is composed of four copies of staggered fermions. Although this additional tensor structure is trivial in the case of the naive fermion, it is nontrivial in the Wilson and minimally doubled fermions and is important when we discuss the symmetry of these fermions in the following sections.

%%%%%%%%%%% Wilson fermion %%%%%%%%%%

\section{Wilson fermion}
\label{sec:WF}
In this section we discuss the symmetry and its breaking in the case of the Wilson fermion with emphasis on the effect of the Wilson term.
As is well-known, the Wilson term splits sixteen doublers into five branches. Since numerical simulations almost exclusively use the ``physical'' branch, which contains only one massless fermion mode, the symmetry and the structure in other branches have not been fully investigated so far.
Therefore, here we will clarify the continuous symmetries and their spontaneous breaking in all the branches. 
As a consequence, we will find an unexpected 
symmetry enhancement and its spontaneous breaking in the central branch.
\subsection{Action and symmetries}
The action for the Wilson fermion~\cite{Wil} is given by
\begin{equation}
S=S_{\mrm{nf}}+S_{W}\quad {\rm with}\quad
S_W=-\frac{r}{2}\sum_{n,\mu}\bar{\psi}_n\left(\psi_{n+\hat{\mu}}-2\psi_n+\psi_{n-\hat{\mu}}\right)\,.
\end{equation}
In terms of the spin-flavor representation,
the Wilson term $S_W$ is written as
\begin{eqnarray}
\nonumber
S_{W}&=&-\frac{r}{2}\sum_{N,\mu}\Big[2\bar{\Psi}(N)\left(\mathbf{1}_4\otimes\gamma_\mu^\T\otimes\gamma_\mu\right)\Psi(N)+\bar{\Psi}(N)\left(\mathbf{1}_4\otimes\gamma_\mu^\T\otimes\gamma_\mu\right)\nabla^2_\mu\Psi(N)\\
\label{WF_action_sfd}
&&+\bar{\Psi}(N)\left(\gamma_\mu\gamma_5\otimes\gamma_5^\T\otimes\gamma_\mu\right)\nabla_\mu\Psi(N)\Big]+ 4 r\sum_{N}\bar{\Psi}(N)\left(\mathbf{1}_4\otimes\mathbf{1}_4\otimes\mathbf{1}_4\right)\Psi(N)\,.
\end{eqnarray}
The first three terms in (\ref{WF_action_sfd}) are invariant under the ordinary $\U(1)$ vector transformation, $\U(1)_V$,
 which is defined by
\begin{eqnarray}
\Psi(N)\rightarrow\Psi^\prime(N)&=&\exp\left[\iu \theta(\mathbf{1}_4\otimes \mathbf{1}_4\otimes \mathbf{1}_4)\right]\Psi(N)\,,\\
\bar{\Psi}(N)\rightarrow\bar{\Psi}^\prime(N)&=&\bar{\Psi}(N)\exp\left[-\iu \theta(\mathbf{1}_4\otimes \mathbf{1}_4\otimes \mathbf{1}_4)\right]\,,
\end{eqnarray}
\begin{equation}
\psi_n\rightarrow\psi_n^\prime=\e^{\iu \theta}\psi_n\,,\quad
\bar{\psi}_n\rightarrow\bar{\psi}_n^\prime=\e^{-\iu \theta}\bar{\psi}_n \, ,
\end{equation}
and the site-dependent $\U(1)$ vector transformation, $\U(1)_V^-$, defined by
\begin{eqnarray}
\Psi(N)\rightarrow\Psi^\prime(N)&=&\exp\left[\iu \theta(\gamma_5\otimes \gamma_5^\T\otimes \mathbf{1}_4)\right]\Psi(N)\,,\\
\bar{\Psi}(N)\rightarrow\bar{\Psi}^\prime(N)&=&\bar{\Psi}(N)\exp\left[\iu \theta(\gamma_5\otimes \gamma_5^\T\otimes \mathbf{1}_4)\right]\,,
\end{eqnarray}
\begin{equation}
\psi_n\rightarrow\psi_n^\prime=\e^{\iu (-1)^{n_1+\ldots +n_4}\theta}\psi_n\,,\quad
\bar{\psi}_n\rightarrow\bar{\psi}_n^\prime=\e^{\iu (-1)^{n_1+\ldots +n_4}\theta}\bar{\psi}_n\,.
\label{U_V^-}
\end{equation}
By contrast the last term in (\ref{WF_action_sfd}) is invariant only under the $\U(1)_V$ transformation.
Therefore, the total Wilson fermion action possesses
only the $\U(1)_V$ symmetry for general values of $m$ and $r$.
Interestingly enough, however, the additional $\U(1)_V^-$ symmetry appears if $m$ and $r$ satisfy $m+4r=0$,
at which the on-site terms cancel out between the mass term and the Wilson term.  As we will show in the next subsection, this symmetry is spontaneously broken by the pion condensate, $\langle \bar{\psi}\gamma_5 \psi\rangle$.

\subsection{Strong coupling analysis}
Now we employ the strong coupling analysis to show that there 
appears an NG boson associated with the $\U(1)_V^-$ symmetry breaking in the presence of the pion condensate. 
An effective action for mesons in the strong coupling limit \cite{KMN1, AokiP, AokiU1} can be written in general as
\begin{eqnarray}
S_{\rm eff}( M) &=& N_c \sum_n \left[\sum_\mu {\rm Tr}\, f(\Lambda_{n,\mu}) + \tr \, \hat M M(n) - \tr \, \log M(n) \right] \, ,\\
\Lambda_{n,\mu}&=&\frac{V_{n,\mu} \bar V_{n,\mu}}{N^2_c} , \quad
M(n)^{\alpha\beta} =  \frac{\sum_a \bar \psi_n^{a,\alpha}\psi_n^{a,\beta}}{N_c}\, ,
\nonumber
\end{eqnarray}
where $N_c$ is the number of colors, ${\rm Tr}$ ( $\tr$ ) means a trace over color(spinor) index, and $M(n)$ is a meson field. The explicit form of the function $f$ is determined by performing a one-link integral of the gauge field. More explicitly we can write
\begin{eqnarray}
V_{n,\mu}^{ab} &=& \bar\psi_n^b P^-_\mu \psi_{n+\hat\mu}^a\, , \quad
\bar V_{n,\mu}^{ab} = -\bar\psi_{n+\hat\mu}^b P^+_\mu \psi_{n}^a\, , \quad \\
{\rm Tr}\, f(\Lambda_{n,\mu})&=& -\tr\, f\left( - M(n) \PpT  M(n+\hat\mu)\PmT
\right)\,, 
\end{eqnarray}
where $4\times 4$ matrices $P_\mu^\pm$ are specified later.
In the large $N_c$ limit, it is known that $f(x)$ can be analytically evaluated as
\begin{equation}
f(x) = \sqrt{1+4x}-1-\ln\frac{1+\sqrt{1+4x}}{2} = x + O(x^2)\, .
\label{eq:largeN}
\end{equation}
However, in the following part of this paper, we will approximate $f(x)$ as $f(x)=x$ unless otherwise stated because qualitative features such as an appearance of NG bosons remain unchanged by this approximation.

To calculate meson masses we expand the meson field as\footnote{In eq. (\ref{sectors}), $S,P, V_{\alpha},A_{\alpha}$ and $T_{\alpha\beta}$ stand for scalar, pseudo-scalar, vector, axial-vector and tensor respectively.} 
\begin{equation}
M(n) = M^\T_0 + \sum_\A \pi^\A(n) \Gamma^\T_\A \,,\quad
\A \in \left\{ S, P, V_\alpha, A_\alpha, T_{\alpha\beta}\right\}\,,
\label{sectors}
\end{equation}
where $M_0$ is the vacuum expectation value (VEV) of $M(n)$, and
\begin{eqnarray}
\Gamma_S =\frac{{\bf 1}_4}{2}, \ \Gamma_P =\frac{\gamma_5}{2}, \ \Gamma_{V_\alpha}=\frac{\gamma_\alpha}{2},  \ \Gamma_{A_\alpha}=\frac{\iu \gamma_5\gamma_\alpha}{2}, \
\Gamma_{T_{\alpha\beta}}=\frac{\gamma_\alpha\gamma_\beta}{2\iu }\ (\alpha < \beta).
\end{eqnarray}
Then the effective action at the second order of $\pi^\A$ is given by
\begin{eqnarray}
S_{\rm eff}^{(2)} &=& N_c \sum_n\biggl[
\frac{1}{2}\tr\,( M_0^{-1} \Gamma_\A M_0^{-1}\Gamma_\B) \,  \pi^\A(n)\pi^\B(n) 
+ \sum_\mu \tr\, ( \Gamma_\A P^{-}_\mu \Gamma_\B P^{+}_\mu ) \pi^\A(n) \pi^\B(n+\hat\mu)
\biggr] \nonumber \\
&=& N_c\int\frac{d^4 p}{(2\pi)^4} \pi^\A(-p) D_{\A\B}(p) \pi^\B(p) \, ,
\end{eqnarray}
where
\begin{eqnarray}
D_{\A\B} (p) &=& \frac{1}{2} \bigl( \widetilde{D}_{\A\B}(p) + \widetilde{D}_{\B\A} (-p)\bigr),\\[.5ex]
\widetilde{D}_{\A\B}(p) &=& \frac{1}{2}\tr\,(M_0^{-1} \Gamma_\A M_0^{-1}\Gamma_\B)
+ \sum_\mu \tr\, ( \Gamma_\A P^-_\mu \Gamma_\B P^+_\mu ) e^{\iu p_\mu} .
\end{eqnarray}

In the case of the Wilson fermion, $\hat M = (m + 4 r)\mathbf{1}_4 \equiv M_W\mathbf{1}_4$ and $P_\mu^\pm =\dfrac{\gamma_\mu\pm r}{2}$.
By taking $M_0=\sigma{\bf 1}_4 + \iu \pi \gamma_5$, we have
\begin{eqnarray}
\left\{
\begin{array}{lll}
\sigma =\dfrac{ - M_W \pm \sqrt{M_W^2+8(1-r^2)}}{4(1-r^2)}\, , & \pi=0 \, ,& M_W^2 \ge M_c^2 \\
\sigma = \dfrac{M_W}{4r^2} \, , & \pi^2 =\dfrac{1}{16r^4(1+r^2)} (8r^4 - M_W^2(1+r^2))\, , & M_W^2 < M_c^2
\end{array}
\right.
\end{eqnarray}
where $M_c^2= \dfrac{8r^4}{1+r^2} $.

\medskip
As discussed in the previous subsection, at $M_W=0$ we have an additional $\U(1)$ symmetry, $\U(1)_V^-$.
Since this parameter regime resides in the parity broken phase, in which $\pi^2\not=0$ and $M_W^2 < M_c^2$, $\U(1)_V^{-}$ is spontaneously broken by the VEV of $\pi$ in this case.

To compute the meson mass, we hereafter take $r^2=1$ for simplicity.
Because $D(p)$ is block-diagonal,
we concentrate on its submatrix $D_{XY}(p)$ with $X, Y \in \left\{ S,P,A_\alpha\right\}$. 
Then, by setting $p=(\pi,\pi,\pi,\pi+\iu  m_{SPA}\,)$,
we find that the $S$-$P$-$A_\alpha$ sector mass $m_{SPA}$ is given by  
\begin{eqnarray}
\cosh (m_{SPA}\,) = 1 + \frac{20 M_W^2}{6-7M_W^2}.
\label{m_SPA}
\end{eqnarray}
Note that since the transformation \eqref{U_V^-} involves the site-dependent quantity $(-1)^{n_1 + \dots + n_4}$, 
it is natural to expand the momentum $p$ around $(\pi,\pi,\pi,\pi)$.
Eq.~\eqref{m_SPA} tells us that the meson becomes a massless NG boson at $M_W=0$ as expected. 
If we use the exact form of $f(x)$ in the large $N_c$ limit, we then obtain
\begin{eqnarray}
\cosh (m_{SPA}\,) = 1 + \frac{2 M_W^2(16+M_W^2)}{16-15M_W^2},
\label{dispersion}
\end{eqnarray}
which again shows that a massless NG boson appears at $M_W=0$.

Before closing this subsection, it is worth noting that the point $M_W=0$ corresponds to the central cusp in the parity broken phase, 
at which six fermion modes with momentum shift, $p=(\pi,\pi,0,0)$,  $(\pi,0,\pi,0)$,  
$(\pi,0,0,\pi)$,  $(0,\pi,\pi,0)$,  $(0,\pi,0,\pi)$ and  $(0,0,\pi,\pi)$, are expected to appear in
the continuum limit.
Although we have not yet known much about the continuum limit for this cusp,  
it is expected to describe QCD with six flavors, which is still asymptotically free.   
Therefore, if an appropriate continuum limit exists, we expect the theory in the limit will be Lorentz-symmetric as in the ``physical" branch because the Wilson fermion action itself possesses 
the hypercubic symmetry.\footnote{
Although the third term in (\ref{WF_action_sfd}) seems to break Lorentz invariance in the continuum limit,
it is just an expressional artifact:
The spin-flavor representation 
does not respect space-time symmetries of the original action~\cite{Lepage}.
Actually such a term cannot exist by itself
because of the original translational invariance.
This representation is not suitable for study of Lorentz symmetry 
although it gives good insight into other symmetries.
}
In addition, the above NG boson propagator recovers the Lorentz-covariant dispersion relation in the naive $a\to0$ limit even in the strong coupling, which is far from the continuum limit.
We also suggest that the Wilson fermion
at $M_{W}=0$ may be applicable to the simulation of the six-flavor QCD without any fine-tunings since the additive mass renormalization is forbidden
by the additional $\U(1)$ symmetry. 
At this point ($M_W=0$), while the chiral condensate $\sigma$ is zero, the pion condensate $\pi$ is non-zero and its magnitude becomes maximal.

%%%%%%%%   Minimally doubled fermion   %%%%%%%%%%%%%

\section{Minimally doubled fermions}
\label{sec:MDF}
Having discussed the naive and the Wilson fermion, now we move on to the analysis on minimally doubled fermions. Minimally doubled fermions are a class of lattice fermions with only two physical fermion modes. Continuous symmetries and their spontaneous breaking for these fermions have not yet been investigated enough. Therefore, in this section, we study two canonical examples of minimally doubled fermions, the Karsten-Wilczek and the Bori\c{c}i-Creutz fermion, with emphasis on symmetry and its spontaneous breaking.
\subsection{Karsten-Wilczek fermion}
\subsubsection{Action and symmetries}
The action for the Karsten-Wilczek fermion \cite{KW} is given by
\begin{equation}
S=S_{\mrm{nf}}+S_\mrm{KW}+S_\mrm{KW}^{(3)}+S_\mrm{KW}^{(4)} \, ,
\end{equation}
with
\begin{eqnarray}
S_\mrm{KW}&=&-\frac{\iu  r}{2}\sum_n\sum_{k=1}^3\bar{\psi}_n\gamma_4\left(\psi_{n+\hat{k}}-2\psi_n+\psi_{n-\hat{k}}\right)\,, \label{SKW1}\\
 S_\mrm{KW}^{(3)}&=&\iu  d_3\sum_n\bar{\psi}_n\gamma_4\psi_n\,,\quad S_\mrm{KW}^{(4)}=\frac{d_4}{2}\sum_{n}\bar{\psi}_n\gamma_4(\psi_{n+\hat{4}}-\psi_{n-\hat{4}})\,,\label{SKW2}
\end{eqnarray}
where $S_\mrm{KW}^{(3)}$ and $S_\mrm{KW}^{(4)}$ are counter terms of dimension
three and four, respectively \cite{CCW}.
This fermion action possesses the cubic subgroup
of the hypercubic symmetry.
We also note it has CT and P invariance while each of 
C and T is broken \cite{BBTW}.

In the free theory,
the Dirac operator in the momentum space becomes
\begin{eqnarray}
D_\mrm{KW}(p) = \iu  \sum_\mu \gamma_\mu \sin p_\mu 
-\iu  r\gamma_4\sum_k (\cos p_k  -1)+ \iu  d_4\gamma_4\sin p_4  + m  + \iu  d_3 \gamma_4,
\end{eqnarray}
whose zero modes appear at $p=(0,0,0,0)$ and $(0,0,0,\pi)$ in the
absence of the mass and counter terms, namely $m  = d_3 = 0$, though
such a fermionic mode can be well-defined only when we
consider the free theory. Note that we  work on the dimensionless action as shown in (\ref{SKW1}) and (\ref{SKW2}); 
thus $r$, $m$, $d_{3}$ and $d_{4}$ all stand for dimensionless parameters.
In the analogy of the Wilson fermion, $d_{3}$ corresponds to the mass parameter $m$ 
because both are parameters for the dimension three terms in the action 
while the parameter $r$ corresponds to the Wilson parameter.
Thus, although the purpose of tuning $d_{3}$ is to recover the hypercubic symmetry, 
it will be tuned in a similar way to the tuning of the mass parameter $m$ 
for the chiral limit in the Wilson fermion in Sec.~\ref{sec:WF}.
On the other hand, $d_{4}$ is also tuned to recover the hypercubic symmetry, 
but there is no corresponding parameter in the Wilson fermion.

\bigskip
We first consider the symmetries of the Karsten-Wilczek term $S_\mrm{KW}$.
In terms of the spin-flavor representation,
$S_\mrm{KW}$ is written as
\begin{eqnarray}
\nonumber
S_\mrm{KW}\hspace{-1mm}&=&\hspace{-1mm}-\frac{\iu  r}{2}\sum_{N,\,k}\Big[2\bar{\Psi}(N)\left(\gamma_4\otimes\gamma_4^\T\gamma_k^\T\otimes\gamma_4\gamma_k\right)\Psi(N)+\bar{\Psi}(N)\left(\gamma_4\otimes\gamma_4^\T\gamma_k^\T\otimes\gamma_4\gamma_k\right)\nabla^2_k\Psi(N)\\
\label{S_KW_sfd}
&&\hspace{-5mm}+\bar{\Psi}(N)\left(\gamma_4\gamma_k\gamma_5\otimes\gamma_4^\T\gamma_5^\T\otimes\gamma_4\gamma_k\right)\nabla_k\Psi(N)\Big]+3\iu  r\sum_{N}\bar{\Psi}(N)\left(\gamma_4\otimes\gamma_4^\T\otimes\gamma_4\right)\Psi(N)\,.
\end{eqnarray}
The first three terms on the right-hand side of~(\ref{S_KW_sfd})
are invariant under the ordinary axial $\U(1)$ symmetry $\U(1)_A$  defined by
\begin{eqnarray}
\Psi(N)\rightarrow\Psi^\prime(N)&=&\exp\left[\iu \theta(\gamma_5\otimes \gamma_5^\T\otimes \gamma_5)\right]\Psi(N)\,,\\
\bar{\Psi}(N)\rightarrow\bar{\Psi}^\prime(N)&=&\bar{\Psi}(N)\exp\left[\iu \theta(\gamma_5\otimes \gamma_5^\T\otimes \gamma_5)\right]\,,
\end{eqnarray}
\begin{equation}
\psi_n\rightarrow\psi_n^\prime=\e^{\iu \theta \gamma_5}\psi_n\,,\quad
\bar{\psi}_n\rightarrow\bar{\psi}_n^\prime=\bar{\psi}_n\e^{\iu \theta\gamma_5}\,,
\end{equation}
and the site dependent axial $\U(1)$ transformation $\U(1)_{A}^+$ \footnote{Here``site dependent'' and ``axial'' refer to original fermion fields $\psi$ and $\bar\psi$.}
defined by
\begin{eqnarray}
\Psi(N)\rightarrow\Psi^\prime(N)&=&\exp\left[\iu \theta(\mathbf{1}_4\otimes \mathbf{1}_4\otimes \gamma_5)\right]\Psi(N)\,,\\
\bar{\Psi}(N)\rightarrow\bar{\Psi}^\prime(N)&=&\bar{\Psi}(N)\exp\left[-\iu \theta(\mathbf{1}_4\otimes \mathbf{1}_4\otimes \gamma_5)\right]\,,
\end{eqnarray}
\begin{equation}
\psi_n\rightarrow\psi_n^\prime=\e^{\iu \theta(-1)^{n_1+\ldots+n_4}\gamma_5}\psi_n\,,\quad
\bar{\psi}_n\rightarrow\bar{\psi}_n^\prime=\e^{-\iu \theta(-1)^{n_1+\ldots+n_4}\gamma_5}\bar{\psi}_n\,,
\end{equation}
in addition to the $\U(1)_V$ and $\U(1)_V^-$, introduced in the previous section.
On the other hand, the last term on the right-hand side of~(\ref{S_KW_sfd}) is invariant only under the $\U(1)_V$ and $\U(1)_A$ transformations.
Therefore the Karsten-Wilczek term $S_\mrm{KW}$ possesses only the ordinary vector and axial-vector $\U(1)$ symmetries, $\U(1)_V$ and $\U(1)_A$.

\medskip
Next let us consider the counter terms $S_\mrm{KW}^{(3)}$ and $S_\mrm{KW}^{(4)}$.
In terms of spin-flavor representation,
they are written as
\begin{eqnarray}
S_\mrm{KW}^{(3)}&=&\iu  d_3\sum_{N}\bar{\Psi}(N)\left(\gamma_4\otimes\gamma_4^\T\otimes\gamma_4\right)\Psi(N)\,,\\
S_\mrm{KW}^{(4)}&=&\frac{d_4}{2}\sum_{N}\Big[\bar{\Psi}(N)\left(\gamma_4\otimes\mathbf{1}_4\otimes\mathbf{1}_4\right)\nabla_4\Psi(N)+\bar{\Psi}(N)\left(\gamma_5\otimes\gamma_4^\T\gamma_5^\T\otimes\mathbf{1}_4\right)\nabla^2_4\Psi(N)\Big]\,.
\end{eqnarray}
We notice that the dimension three counter term $S_\mrm{KW}^{(3)}$
takes the same form as the last term on the right-hand side of~(\ref{S_KW_sfd}). Therefore, the dimension three counter term is invariant under $\U(1)_V$ and $\U(1)_A$.
On the other hand, the dimension four counter term $S_\mrm{KW}^{(4)}$
is invariant under all of the $\U(4)\times \U(4)$ transformations.

\medskip
From the above discussions,
the Karsten-Wilczek fermion action
possesses only the $\U(1)_V$ symmetry for general values of $m$, $r$, $d_3$ and $d_4$,
and it acquires the $\U(1)_A$ symmetry at $m=0$.
If $d_3+3r=0$ and $m=0$ are simultaneously satisfied,
it further acquires the $\U(1)_V^-$ and the $\U(1)_A^+$ symmetry
in addition to $\U(1)_V$ and $\U(1)_A$.

\subsubsection{Strong coupling analysis}
\label{sec:KW}
In the case of the Karsten-Wilczek fermion, we have $\hat M = m {\bf
1}_4 + \iu (d_3+3r)\gamma^\T_4$ and
\begin{eqnarray}
P^+_\mu &=& \left\{
\begin{array}{ccc}
 \frac{1}{2} (\gamma_\mu + \iu  r \gamma_4) & \mu=1,2,3     \\
 \frac{1}{2}\gamma_4(1+d_4) & \mu=4   \\
\end{array}
\right. ,
 \quad
P^-_\mu =
 \left\{
\begin{array}{ccc}
 \frac{1}{2} (\gamma_\mu - \iu  r \gamma_4) & \mu=1,2,3     \\
 \frac{1}{2}\gamma_4(1+d_4) & \mu=4   \\
\end{array}
\right. .
\end{eqnarray}
As we discussed in the previous subsection,
$r$ corresponds to the Wilson parameter while $d_{3}$ corresponds to
the mass parameter $m$ in the analogy of Wilson fermion.
Thus, in the following, $d_3$ will be tuned as the mass parameter 
$m$ in the Wilson fermion was tuned as $m+4r=0$ in Sec.~\ref{sec:WF}.

By taking $M_0 =\sigma{\bf 1}_4 + \iu \pi_4\gamma_4$, the corresponding gap equations become
\begin{eqnarray}
\frac{3(1+r^2)+(1+d_4)^2}{2}\sigma + m -\frac{\sigma}{\sigma^2+\pi_4^2} &=& 0 \, ,\\
\frac{3(1-r^2)-(1+d_4)^2}{2}\pi_4 -(d_3+3r) -\frac{\pi_4}{\sigma^2+\pi_4^2} &=& 0 \,.
\end{eqnarray}
Since the VEV of $\pi_4$ would violate the Lorentz invariance, we would
like to consider a solution with $\sigma\not=0$ and $\pi_4=0$. To obtain
a solution with $\pi_4=0$, we need to tune $d_3+3r =0$ for the dimension
three counter term. 
In this case we need to impose
\begin{eqnarray}
\frac{2}{\sigma^2} = \frac{2m}{\sigma}+3(1+r^2)+(1+d_4)^2 .
\end{eqnarray}  
As discussed in the last subsection, at $m=0$ and $d_3+3r=0$, the action is invariant under an additional $\U(1)_V^-$ and $\U(1)_A^+$ symmetries in addition to $\U(1)_V$ and $\U(1)_A$.
Since the chiral condensate $\langle \bar\psi_n\psi_n\rangle \not=0$ spontaneously breaks
$\U(1)_A$ and $\U(1)_V^-$,
we expect two massless NG bosons at $m=0$.
Note that the number of fermion zero modes at $d_3+3r=0$ is not two but does depend on the values of $r$ and $d_4$. At $r=1$ and $d_4=0$, for example, there appear six zero modes at $p=(0,0,\pi,\pi/2)$,  $(0,\pi,0,\pi/2)$,  $(\pi,0,0,\pi/2)$,  $(0,\pi,\pi,-\pi/2)$,  $(\pi,\pi,0,-\pi/2)$ and  $(\pi,0,\pi,-\pi/2)$.

The inverse meson propagator matrix for $S$-$P$-$T_{\alpha\beta}$ sector can be factorized as
\begin{eqnarray}
D^{SPT} &=& 
\left(
\begin{array}{cc}
D^{ST} & 0 \\
0 & D^{PT} \\
\end{array}
\right) \, ,
\end{eqnarray}
where the matrix $D^{ST}$ is defined by  
\begin{eqnarray}
D^{ST}_{XY}(p) &=& 
\left(
\begin{array}{cc}
\delta_{ab} D_a(p) & -\iu \frac{r}{2}s_a     \\
  \iu \frac{r}{2} s_b & D_S(p)     \\
\end{array}
\right) \, ,\\
D_a(p) &=& \frac{1}{2\sigma^2}+\frac{1}{4}\left[ \sum_k \{(1+r^2) -2\delta_{ka} -2r^2\}c_k-(1+d_4)^2c_4\right] \, ,\\
D_S(p) &=&  \frac{1}{2\sigma^2}+\frac{1}{4}\left[ \sum_k (1+r^2)c_k +(1+d_4)^2c_4\right] \, ,
\end{eqnarray}
for $ X =(T_{a4}, S)$ and $Y=(T_{b4},S)$, and the matrix $D^{PT}$ is defined by 
\begin{eqnarray}
D^{PT}_{XY}(p) &=& 
\left(
\begin{array}{cc}
\delta_{ac}\delta_{bd} D_{ab}(p) & -\iu \varepsilon^{abe}\frac{r}{2}s_e     \\
  \iu \frac{r}{2}\varepsilon^{cde} s_e & D_P(p)   \\
\end{array}
\right) \, ,\\
D_{ab}(p) &=& \frac{1}{2\sigma^2}+\frac{1}{4}\left[ \sum_k \{(1+r^2) -2\delta_{ka} -2\delta_{kb}\}c_k+(1+d_4)^2c_4\right] \, ,\\
D_P(p) &=&  \frac{1}{2\sigma^2}-\frac{1}{4}\left[ \sum_k (1+r^2)c_k +(1+d_4)^2c_4\right] \, ,
\end{eqnarray}
for $ X =(T_{ab}, P)$ and $Y=(T_{cd},P)$.  Here $s_a =\sin(p_a)$ and $c_a=\cos(p_a)$.
Since $D^{ST}(\pi+p) = D^{PT}( p)$, it is enough to first consider $D^{PT}(p)$ only and then double the degeneracy of the spectrum.

By taking $p=(0,0,0,\iu  m_P)$, we obtain
\begin{eqnarray}
\cosh(m_P) &=& 1+\frac{1}{(1+d_4)^2}\frac{2m}{\sigma}\, .
\end{eqnarray}
This mode corresponds to a massless NG boson at $m=0$. Together with the corresponding mode in the $S$ sector, totally two NG bosons appear at $m=0$ as expected. 

If we introduce a small but non-zero spatial momentum, energy of the meson in the $P-T$ sector becomes $E^2 = m_P^2 + X (\vec p)^2$ where
\begin{eqnarray}
X&=&\frac{m_{P}}{\sinh(m_{P})}\times \frac{1}{(1+d_4)^2}\left[(1+r^2) -\frac{2r^2}{T}\right] \, ,\\
T&=&  \frac{m}{2\sigma}+\frac{1}{2}\{1+3r^2+(1+d_4)^2\}\, .
\end{eqnarray} 
To require the Lorentz covariance for the dispersion relation at $m=0$, we need to choose $d_4$ as
\begin{eqnarray}
(1+d_4)^2 &=& -r^2 + \sqrt{1+4r^4} \, .
\end{eqnarray}
If we use the exact form of $f(x)$ in the large $N_c$ limit, we then obtain
\begin{eqnarray}
&&\cosh m_P=1+\frac{m}{A\sigma}\, ,\\
&&A=\frac{(1+d_4)^2}{4\sqrt{1-\sigma^2 (1+d_4)^2}}\, ,
\end{eqnarray}
which again corresponds to a massless Goldstone boson at $m=0$. The introduction of small but finite spatial momenta results in
\begin{eqnarray}
&&E^2=m_P^2 +X(\vec{p})^2\, ,\\
&&X=\frac{m_P}{A\sinh m_P}\left( B-\frac{2B^2}{T^{\prime}}\right)\, ,\\
&&T^{\prime}=2\left( A+\frac{m}{\sigma}\right) +4(B-C)\, ,\\
&&B=\frac{1}{2\sqrt{1-4\sigma^2}}\, ,\\
&&C=\frac{\sigma ^2}{(1+\sqrt{1-4\sigma ^2})^2\sqrt{1-4\sigma ^2}}\, .
\end{eqnarray}
This shows that we can recover the relativistic dispersion by tuning $d_4$. As the discussion here is only on the strong coupling limit, the above results are not directly related to the properties of the continuum limit. However, the strong coupling analysis in this subsection exhibits several important features, such as the rotational symmetry breaking and its restoration by tuning counter terms, which are expected to appear also in the study of the continuum limit. We thus consider that our study is useful, at least qualitatively, for the investigation of the continuum limit of the minimally doubled fermions at $d_3+3r=0$. 

\subsection{Bori\c{c}i-Creutz fermion}
\subsubsection{Action and symmetries}
The action for the Bori\c{c}i-Creutz  fermion \cite{CB} is given by
\begin{equation}
S=S_{\mrm{nf}}+S_\mrm{BC}+S_\mrm{BC}^{(3)}+S_\mrm{BC}^{(4)} \, ,
\end{equation}
with
\begin{eqnarray}
S_\mrm{BC}&=&\frac{\iu  r}{2}\sum_{n,\,\mu}\bar{\psi}_n(\Gamma-\gamma_\mu)\left(\psi_{n+\hat{\mu}}-2\psi_n+\psi_{n-\hat{\mu}}\right)\,,\\
 S_\mrm{BC}^{(3)}&=&\iu  c_3\sum_n\bar{\psi}_n\Gamma\psi_n\,,\quad S_\mrm{BC}^{(4)}=\frac{c_4}{2}\sum_{n,\,\mu}\bar{\psi}_n\Gamma(\psi_{n+\hat{\mu}}-\psi_{n-\hat{\mu}})\,,\\
\Gamma &=&\frac{1}{2}\sum_\mu \gamma_\mu, \quad \gamma_\mu^\prime =\Gamma\gamma_\mu\Gamma =\Gamma-\gamma_\mu,
\end{eqnarray}
where $S_\mrm{BC}^{(3)}$ and $S_\mrm{BC}^{(4)}$ are counter terms of dimension three and four, respectively \cite{CCWW, Cichy}, and $\Gamma$ satisfies $\Gamma^2=1$ and $\{ \Gamma,\gamma_\mu\} = 1$. This action has the ${\rm S}_{4}$ subgroup of the hypercubic symmetry while
C, P and T symmetries are broken to the combined CPT \cite{BBTW}. 

In the free theory, the Dirac operator in the momentum space can be expressed as
\begin{eqnarray}
D_\mrm{BC}(p) = \iu  \sum_\mu\left[ \gamma_\mu \sin p_\mu  + r \gamma_\mu^\prime (\cos p_\mu -1) +c_4\Gamma\sin p_\mu \right] + m + \iu  c_3 \Gamma.
\end{eqnarray}
If we consider the case with $m = c_3 = 0$,
$D_\mrm{BC}(p)$ has one zero mode at $p= (0,0,0,0)$, and another one at $p =(\pi/2,\pi/2,\pi/2,\pi/2)$ if
\begin{eqnarray}
\sum_\mu (\gamma_\mu + c_4 \Gamma -r\gamma_\mu^\prime )&=&
2\Gamma + 4 c_4\Gamma -2 r \Gamma = 0,
\end{eqnarray}
which determines $c_4 =(r-1)/2$. 
We note there are generically two zero modes in the momentum space, 
but their positions depend on the choice of parameters.
As with the Karsten-Wilczek fermion,
$c_{3}$ corresponds to $m$ while $r$ corresponds to the Wilson parameter
in the analogy of the Wilson fermion.
Thus $c_{3}$ will be tuned as $m$ for the Wilson fermion and $d_{3}$ for
the Karsten-Wilczek fermion.

\bigskip
First we consider the symmetries of the Bori\c{c}i-Creutz term $S_\mrm{BC}$.
In terms of the spin-flavor representation,
it is written as
\begin{eqnarray}
\nonumber
S_\mrm{BC}&=&\frac{\iu  r}{4}\sum_{N,\,\mu,\,\nu}\Big[2\bar{\Psi}(N)\left(\gamma_\nu\otimes\gamma_\nu^\T\gamma_\mu^\T\otimes\gamma_\nu\gamma_\mu\right)\Psi(N)+\bar{\Psi}(N)\left(\gamma_\nu\otimes\gamma_\nu^\T\gamma_\mu^\T\otimes\gamma_\nu\gamma_\mu\right)\nabla_\mu^2\Psi(N)
\\
\nonumber
&&\hspace{15mm}+\bar{\Psi}(N)\left(\gamma_\nu\gamma_\mu\gamma_5\otimes\gamma_\nu^\T\gamma_5^\T\otimes\gamma_\nu\gamma_\mu\right)\nabla_\mu\Psi(N)\Big]\\
\nonumber
&&-\frac{\iu  r}{2}\sum_{N,\,\mu}\Big[2\bar{\Psi}(N)\left(\gamma_\mu\otimes\mathbf{1}_4\otimes\mathbf{1}_4\right)\Psi(N)+\bar{\Psi}(N)\left(\gamma_\mu\otimes\mathbf{1}_4\otimes\mathbf{1}_4\right)\nabla_\mu^2\Psi(N)
\\
\nonumber
&&\hspace{15mm}+\bar{\Psi}(N)\left(\gamma_5\otimes\gamma_\mu^\T\gamma_5^\T\otimes\mathbf{1}_4\right)\nabla_\mu\Psi(N)\Big]\\
\label{S_BC_sfd}
&&-\iu  r\sum_{N,\mu}\bar{\Psi}(N)\left(\gamma_\mu\otimes\gamma_\mu^\T\otimes\gamma_\mu\right)\Psi(N)\,.
\end{eqnarray}
While the first six terms on the right-hand side of~(\ref{S_BC_sfd})
are invariant under $\U(1)_V$, $\U(1)_A$, $\U(1)_V^-$, and $\U(1)_A^+$,
the last term is invariant only under $\U(1)_V^-$ and $\U(1)_A^+$.
Therefore the Bori\c{c}i-Creutz term $S_\mrm{BC}$ possesses only the ordinary vector and  axial-vector $\U(1)$ symmetry, $\U(1)_V$ and $\U(1)_A$.

\medskip
In terms of spin-flavor representation, the counter terms $S_\mrm{BC}^{(3)}$ and $S_\mrm{BC}^{(4)}$ are written as
\begin{eqnarray}
S_\mrm{BC}^{(3)}&=&\frac{\iu  c_3}{2}\sum_{N,\mu}\bar{\Psi}(N)\left(\gamma_\mu\otimes\gamma_\mu^\T\otimes\gamma_\mu\right)\Psi(N)\, ,\\
\nonumber
S_\mrm{BC}^{(4)}&=&\frac{c_4}{4}\sum_{N,\,\mu,\,\nu}\Big[\bar{\Psi}(N)\left(\gamma_\nu\otimes\gamma_\nu^\T\gamma_\mu^\T\otimes\gamma_\nu\gamma_\mu\right)\nabla_\mu\Psi(N)\\
&&\hspace{16mm}+\bar{\Psi}(N)\left(\gamma_\nu\gamma_\mu\gamma_5\otimes\gamma_\nu^\T\gamma_5^\T\otimes\gamma_\nu\gamma_\mu\right)\nabla^2_k\Psi(N)\Big]\,.
\end{eqnarray}
We notice that the dimension three counter term $S_\mrm{BC}^{(3)}$
takes the same form as the last term on the right-hand side of~(\ref{S_BC_sfd}). Therefore, the dimension three counter term is invariant only under $\U(1)_V$ and $\U(1)_A$. 
On the other hand, the dimension four counter term, $S_\mrm{BC}^{(4)}$,
is invariant under $\U(1)_V$, $\U(1)_A$, $\U(1)_V^-$, and $\U(1)_A^+$.

\medskip
From the above discussions,
the Bori\c{c}i-Creutz fermion action
possesses only the $\U(1)_V$ symmetry for general values of $m$, $r$, $c_3$ and $c_4$,
and it acquires the $\U(1)_A$ symmetry at $m=0$.
If $c_3-2r=0$ is satisfied at the same time,
it possesses $\U(1)_V^-$ and $\U(1)_A^+$
in addition to the ordinary vector and axial-vector $\U(1)$ symmetries, $\U(1)_V$ and $\U(1)_A$.

\subsubsection{Strong coupling analysis}
For the Bori\c{c}i-Creutz fermion we have $\hat M = m{\bf 1}_4 + \iu (c_3-2r)\Gamma^\T$ and
\begin{eqnarray}
P^+_\mu &=& \frac{1}{2} \{\gamma_\mu (1+\iu  r) +(c_4-\iu  r)\Gamma \}, \quad
P^-_\mu = \frac{1}{2} \{\gamma_\mu (1-\iu  r) +(c_4+\iu  r)\Gamma \} .
\end{eqnarray}
We note $c_{3}$ will be tuned as with $d_{3}$ for the Karsten-Wilczek fermion.
By taking $M_0 = \sigma{\bf 1}_4 + \iu \pi_\Gamma \Gamma$, the corresponding gap equations become
\begin{eqnarray}
 2(1+r^2+c_4^2+c_4) \sigma +m - \frac{\sigma}{\sigma^2+\pi_\Gamma^2} &=& 0 \, ,\\
 (1+r^2 -2c_4^2-2c_4)\pi_\Gamma -(c_3-2 r) - \frac{\pi_\Gamma}{\sigma^2+\pi_\Gamma^2} &=& 0 \, , 
\end{eqnarray}
To have $\pi_\Gamma =0$, we need $c_3-2r = 0$. In this case, $\sigma$ is determined by
\begin{eqnarray}
\frac{1}{\sigma^2} &=& \frac{m}{\sigma} + 2(1+r^2+c_4^2+c_4) .
\end{eqnarray}
As discussed in the last subsection,
at $m=0$ and $c_3-2r=0$, the action is invariant under the $\U(1)_V^-$ and the $\U(1)_A^+$ symmetries in addition to $\U(1)_V$ and $\U(1)_A$.
Since the chiral condensate $\langle \bar\psi_n\psi_n\rangle \not=0$ breaks
$\U(1)_A$ and $\U(1)_V^-$,
we expect two massless NG bosons at $m=0$. Note that the number of fermion zero modes at $c_3-2r=0$ depends on $r$ and $c_4$. For example, at $r=1$ and $c_4=-1$, there are sixteen zero modes at $p_\mu = \pi/2$ or $\pi$ for each $\mu$.

In the strong coupling limit,  there exists non-trivial mixing between mesons, which complicates the analysis on the meson mass. Fortunately, by taking $c_4+1=0$, we can avoid such mixing. For this choice,
the inverse propagator for the $S$ or $P$ sector can be expressed~as
\begin{eqnarray}
D_{SS}(p) &=&  \frac{1}{2\sigma^2}+\frac{1+r^2}{4}\sum_\mu \cos p_\mu, \\
D_{PP}(p) &=&  \frac{1}{2\sigma^2}-\frac{1+r^2}{4}\sum_\mu \cos p_\mu.
\end{eqnarray}
Note that to derive the above expressions we have already used $c_4+1=0$.
By solving $D_{XX}(p+\Delta_X \pi )=0$ for $X=S$ or $P$ with $p=(\vec p, E_X)$, $\Delta_S^{\mu}=1$ and $\Delta_P^{\mu}=0$, we obtain
\begin{eqnarray}
2(\cosh(E_X) -1) &=& \frac{2m}{\sigma} \frac{2}{1+r^2} +(\vec p)^2,
\end{eqnarray}
which has the relativistic dispersion relation at $m=0$. These two modes correspond to NG bosons associated with the breakdown of two $\U(1)$ symmetries as expected.

%%%%%%%%%%   Summary and Discussion   %%%%%%%%%%

\section{Summary and Discussion}
\label{sec:SD}
In this paper we clarify the detailed structure of continuous symmetry 
and its spontaneous breaking in four types of lattice fermion using the 
spin-flavor representation and the strong coupling analysis. 

We begin by reviewing the $\U(4)\times\U(4)$ symmetries of the naive fermion from the viewpoint of the spin-flavor representation. These symmetries can be interpreted 
as generalizations of $\U(1)\times \U(1)$  symmetry of 
the staggered fermion and are a subgroup of the $\U(16)\times\U(16)$ 
symmetries, which are expected to be restored in the continuum 
limit. 
We then apply the same method to 
the Wilson fermion and discover an incidental symmetry enhancement in the 
central branch of the Dirac spectrum 
(or equivalently, at the third cusp in the parity-broken phase). In this branch, on-site terms 
cancel out between the Wilson term and the mass term and an extra $\U(1)$ symmetry 
emerges in addition to the ordinary vector $\U(1)$ symmetry. Using the strong 
coupling analysis, we show that this extra symmetry is spontaneously broken 
by the pion condensate.
We also discuss two types of minimally doubled fermions, the Karsten-Wilczek 
and the Bori\c{c}i-Creutz fermions.  The symmetry enhancement and its 
breaking also occur in these fermions when on-site terms are absent.  
Enhanced symmetries are spontaneously broken by the ordinary chiral condensate in this case.

\bigskip
Now let us discuss possible applications of our results. 
%Firstly, our analysis on the detailed structure of symmetry in the naive fermion may be useful
%to simulate sixteen flavor QCD with the naive fermions, which is believed to be inside the 
%conformal window where the asymptotic freedom and the IR fixed point coexist. 
%The detailed knowledge on symmetry will greatly 
%alleviate the difficulty of parameter tuning in the simulation.
%Secondly,
First,
we suggest that the central branch with six species of doublers in the 
Wilson fermion is potentially useful for the simulation of many flavor QCD, 
especially for the simulation with six or twelve flavors. As elucidated in this 
paper, the action of the Wilson fermion without on-site terms has the enhanced 
$\U(1)$ symmetry. Since this symmetry will prohibit appearance of mass terms 
through quantum correction, we are free from additive mass renormalization in 
this case. 
In this branch, the roles of $\sigma$ and $\pi$ mesons are interchanged. The enhanced 
$\U(1)$ symmetry, which is broken by the pion condensate, is 
considered as a counterpart of the flavor non-singlet chiral symmetry. 
Although this $\U(1)$ is expected to enhance to $\U(6)\times \U(6)$ 
in the continuum limit of the free theory, it is not clear at present whether 
this flavor symmetry is intact also in the presence of gauge fields.
If we could take a continuum limit of the Wilson fermion at the central 
cusp which restores the flavor symmetry,
it would be valuable since it would give an alternative approach to the twelve flavor QCD:
there is still a controversy on 
whether this theory is in the conformal window or not~\cite{Appel, Deuz, Onogi, Hasen, Hayakawa, DeGrand, Debbio}.
%It is quite valuable if we can simulate 
%QCD with a multiple of six flavors in this way, since there is still a controversy on 
%whether QCD with twelve flavors is in the conformal window or not~\cite{Appel, Deuz, Onogi, Hasen, Hayakawa, DeGrand, Debbio}.

%Finally,
Second,
we comment on the implication of our results on simulations with 
minimally doubled fermions. When one uses minimally 
doubled fermions, one needs to fine-tune several parameters to restore 
the Lorentz covariance, broken by the actions, 
 in the continuum limit. To clarify this parameter tuning process and 
restoration of the Lorentz covariance, several works on the nonperturbative 
renormalization have already been done \cite{CCW,CCWW,Cichy}. However restorations 
of the Lorentz covariance in the absence of on-site terms, discussed
in this paper,  is not directly relevant to the nonperturbative 
renormalization of {\it minimally doubled} fermions, since a number of fermion 
zero modes is actually {\it not} minimal without on-site terms. To study the 
non-perturbative renormalization of minimally doubled fermions by the 
strong-coupling analysis, we need to find other tuning points at which the 
number of fermion zero modes is truly minimal. At such points, the condensate 
$\langle \bar\psi \gamma_4 \psi\rangle$ or  $\langle \bar\psi \Gamma \psi\rangle$ 
is generally  nonzero. Investigations into such parameter regions 
are currently in progress.

%%%%%%%%%%   ACKNOWLEDGMENTS   %%%%%%%%%%
\bigskip
\noindent
{\bf \large Acknowledgments}

\medskip
SA would like to thank Institute of Physics, University of Tokyo (Komaba) for 
kind hospitalities during his lectures at the graduate course,  which have brought 
us this research project. This work is supported in part by the Grant-in-Aid for 
Scientific Research on Innovative Areas (No. 2004:  20105001, 20105003).
TM appreciates fruitful discussion with Michael Creutz and Taku Izubuchi.
TK, SK, TM, TN, and ST are supported in part by Grand-in-Aid for the Japan 
Society for Promotion of Science (JSPS) Research Fellows.

%%%%%%%%%%   APPENDIX   %%%%%%%%%%

\appendix

%%%%%%%%%Appendix A%%%%%%%%%%
\section{Spin-flavor representation of fermion actions}
In this appendix, we present some details on how we rewrite fermion actions in terms of the spin-flavor representation.

Let us first describe the case of naive fermions. The action of the naive fermion,  expressed by $\chi$ as in (\ref{spin-diagonalized_action}), can be recast in the following form.
\begin{equation}
\begin{split}
S_{\mrm{nf}}=&\frac{1}{2}\sum_{N,\,\mu,\,A,\,B}\eta_\mu(A)\bar{\chi}_A(N)\Big((\delta_{A+\hat{\mu},\,B}+\delta_{A-\hat{\mu},\,B})\nabla_\mu\chi_B(N)-(\delta_{A+\hat{\mu},\,B}-\delta_{A-\hat{\mu},\,B})\nabla_\mu^2\chi_B(N)\Big)\\
&+m\sum_{N,\,A}\bar{\chi}_A(N)\chi_A(N)\,,
\end{split}
\end{equation}
where $A_{\mu}$ and $B_{\mu}$ take zero or one, and the definitions of $\nabla_{\mu}$ and $\nabla_{\mu}^2$ are given in (\ref{nabla}) and (\ref{nabla2}). In order to further rewrite the above action using the field $\Psi$, defined in (\ref{Psi}) and (\ref{Psibar}), the following formulas are useful;
\begin{equation}
\begin{split}
&\sum_{A,\,B}\eta_\mu(A)(\delta_{A+\hat{\mu},\,B}+\delta_{A-\hat{\mu},\,B})\left(\frac{\gamma_A}{2}\right)_{\alpha,\,f_1}\left(\frac{\bar{\gamma}_B}{2}\right)_{\alpha^\prime,\,f^\prime_1}=\sum_{A}\left(\frac{\gamma_A}{2}\right)_{\alpha,\,f_1}\left(\gamma^\T_\mu\,\frac{\bar{\gamma}_{A}}{2}\right)_{\alpha^\prime,\,f^\prime_1}
\\
&=(\gamma_\mu)_{\alpha,\,\alpha^\prime}\,\delta_{f_1,\,f_1^\prime} \, ,\label{fml1}
\end{split}
\end{equation}
and
\begin{equation}
\begin{split}
&\sum_{A,\,B}\eta_\mu(A)(\delta_{A+\hat{\mu},\,B}-\delta_{A-\hat{\mu},\,B})\left(\frac{\gamma_A}{2}\right)_{\alpha,\,f_1}\left(\frac{\bar{\gamma}_B}{2}\right)_{\alpha^\prime,\,f^\prime_1}=\sum_{A}\left(\gamma_5\,\frac{\gamma_A}{2}\,\gamma_5\right)_{\alpha,\,f_1}\left(\frac{\bar{\gamma}_{A}}{2}\,\gamma^\T_\mu\right)_{\alpha^\prime,\,f^\prime_1}\\
&=(\gamma_5)_{\alpha,\,\alpha^\prime}\,(\gamma_5^\T\gamma_\mu^\T)_{f_1,\,f_1^\prime}\,. 
\end{split}
\label{fml2}
\end{equation}
To derive (\ref{fml1}) and (\ref{fml2}), it is convenient to first consider the cases when $A_{\mu}=0$ and $A_{\mu}=1$ separately and put them together at the end. Using these formulas one obtains the expression (\ref{naive_sfd}) for the naive fermion.

In the case of Wilson fermions, we also need to re-express the Wilson term, which can be represented in terms of $\chi_A(N)$ as
\begin{equation}
\begin{split}
S_W=&-\frac{r}{2}\sum_{N,\,\mu,\,A,\,B}\tilde{\eta}_\mu(A)\bar{\chi}_A(N)\gamma_\mu
\Big[2(\delta_{A+\hat{\mu},\,B}+\delta_{A-\hat{\mu},\,B})\chi_B(N)-(\delta_{A+\hat{\mu},\,B}-\delta_{A-\hat{\mu},\,B})\nabla_\mu\chi_B(N)\\
&+(\delta_{A+\hat{\mu},\,B}+\delta_{A-\hat{\mu},\,B})\nabla_\mu^2\chi_B(N)\Big]
+4r\sum_{N,A}\bar{\chi}_A(N)\chi_A(N)\,.
\end{split}
\end{equation}
To derive the expression in terms of $\Psi$, this time one needs the variants of (\ref{fml1}) and (\ref{fml2});
\begin{equation}
\begin{split}
&\sum_{A,\,B}\tilde{\eta}_\mu(A)(\delta_{A+\hat{\mu},\,B}+\delta_{A-\hat{\mu},\,B})\left(\frac{\gamma_A}{2}\right)_{\alpha,\,f_1}\left(\frac{\bar{\gamma}_B}{2}\right)_{\alpha^\prime,\,f^\prime_1}=\sum_{A}\left(\frac{\gamma_A}{2}\right)_{\alpha,\,f_1}\left(\frac{\bar{\gamma}_{A}}{2}\,\gamma^\T_\mu\right)_{\alpha^\prime,\,f^\prime_1}\\
&=\delta_{\alpha,\,\alpha^\prime}\,(\gamma_\mu^\T)_{f_1,\,f_1^\prime}\, ,
\end{split}
\end{equation}
and
\begin{equation}
\begin{split}
&\sum_{A,\,B}\tilde{\eta}_\mu(A)(\delta_{A+\hat{\mu},\,B}-\delta_{A-\hat{\mu},\,B})\left(\frac{\gamma_A}{2}\right)_{\alpha,\,f_1}\left(\frac{\bar{\gamma}_B}{2}\right)_{\alpha^\prime,\,f^\prime_1}
=\sum_{A}\left(\gamma_5\,\frac{\gamma_A}{2}\,\gamma_5\right)_{\alpha,\,f_1}\left(\gamma^\T_\mu\,\frac{\bar{\gamma}_{A}}{2}\right)_{\alpha^\prime,\,f^\prime_1}\\
&=(\gamma_5\gamma_\mu)_{\alpha,\,\alpha^\prime}\,(\gamma_5^\T)_{f_1,\,f_1^\prime}\,.
\end{split}
\end{equation}
Applying these formulas one can re-express the Wilson term and obtain (\ref{WF_action_sfd}).

The actions of minimally doubled fermions can also be re-expressed by $\Psi$ in a similar way and the resultant expressions are (\ref{S_KW_sfd}) and (\ref{S_BC_sfd}). 
%%%%%%%%%Appendix B%%%%%%%%%%%

\section{Strong coupling analysis for Bori\c{c}i-Creutz fermion }
In this appendix, some details of our analysis in the strong coupling limit for Bori\c{c}i-Creutz fermion are given.

Before calculating the inverse propagator, it is useful to consider the following trace formulas. 
\begin{eqnarray}
\tr\, ( \Gamma_{V_a} P^+_\mu \Gamma_{V_b} P^-_\mu ) &=&
- \tr\, ( \Gamma_{A_a} P^+_\mu \Gamma_{A_b} P^-_\mu ) =\frac{\delta_{ab}}{4}\left\{(1+r^2)(2\delta_{a\mu}-1)-c_4^2-c_4\right\} \nonumber \\
&+&\frac{c_4^2+r^2}{8}+\frac{c_4-r^2}{4}(\delta_{a\mu}+\delta_{b\mu}) 
\, ,\\
\tr\, ( \Gamma_{V_a} P^+_\mu \Gamma_{A_b} P^-_\mu ) &=&
- \tr\, ( \Gamma_{A_b} P^+_\mu \Gamma_{V_a} P^-_\mu ) 
= \frac{r(1+c_4)}{4}\sum_{\nu}\epsilon ^{ab\mu\nu}  \, ,
\label{eq:VAmixing}\\
\tr\, ( \Gamma_{S} P^+_\mu \Gamma_{S} P^-_\mu ) &=&
-\tr\, ( \Gamma_{P} P^+_\mu \Gamma_{P} P^-_\mu ) =
\frac{1+r^2+c_4^2+c_4}{4} 
\, ,\\
\tr\, ( \Gamma_{T_{ab}} P^+_\mu \Gamma_{T_{cd}} P^-_\mu ) &=&
\delta_{ac}\delta_{bd}\frac{(1+r^2)(1-2\delta_{a\mu}-2\delta_{b\mu}) +(c_4^2+r^2)+(c_4-r^2)}{4} \nonumber \\
&+&\frac{c_4^2+r^2}{8}(\delta_{ad}+\delta_{bc}-\delta_{bd}-\delta_{ac}) 
+\frac{c_4-r^2}{4}\left\{(\delta_{a\mu}(\delta_{bc}-\delta_{bd}) +\delta_{b\mu}(\delta_{ad}-\delta_{ac})\right.\nonumber \\
&+& \left. \delta_{c\mu}(\delta_{ad}-\delta_{bd})+\delta_{d\mu}(\delta_{bc}-\delta_{ac})   \right\} 
\, ,\\
\tr\, ( \Gamma_{S} P^+_\mu \Gamma_{T_{ab}} P^-_\mu ) &=&
-\tr\, ( \Gamma_{T_{ab}} P^+_\mu \Gamma_{S} P^-_\mu ) = \frac{r(1+c_4)}{4}(\delta_{a\mu}-\delta_{b\mu}) 
\, ,\label{eq:STmixing}\\
\tr\, ( \Gamma_{P} P^+_\mu \Gamma_{T_{ab}} P^-_\mu ) &=&
-\tr\, ( \Gamma_{T_{ab}} P^+_\mu \Gamma_{P} P^-_\mu ) = \frac{r(1+c_4)}{4}\sum_\nu \varepsilon^{ab\mu\nu} \, .
\label{eq:PTmixing}
\end{eqnarray}
Since eqs.~(\ref{eq:VAmixing}), (\ref{eq:STmixing}) and (\ref{eq:PTmixing}) produce mixing among several sectors, which complicates our analysis on the meson mass,  
we tune $c_4$ in order to avoid such mixings: $c_4+1=0$.

Then, an inverse propagator for  the $S$ or $P$ sector is given by 
\begin{eqnarray}
D_S(p) &=&  \frac{1}{2\sigma^2}+\frac{1+r^2}{4}\sum_\mu \cos p_\mu, \\
D_P(p) &=&  \frac{1}{2\sigma^2}-\frac{1+r^2}{4}\sum_\mu \cos p_\mu,
\end{eqnarray}
where we have already used $c_4+1=0$.

An inverse propagator matrix for $V$ or $A$ sector is given by
\begin{eqnarray}
D_{V_aV_b}(p) &=&
\delta_{ab}\left[ \frac{1}{2\sigma^2} +\frac{1+r^2}{4}(2c_a -\sum_\mu c_\mu)\right]
+\frac{1+r^2}{8}\sum_\mu c_\mu -\frac{1+r^2}{4}(c_a+c_b) \, ,\\
D_{A_aA_b}(p) &=&
\delta_{ab}\left[ \frac{1}{2\sigma^2} -\frac{1+r^2}{4}(2c_a -\sum_\mu c_\mu)\right]
-\frac{1+r^2}{8}\sum_\mu c_\mu +\frac{1+r^2}{4}(c_a+c_b)\, ,
\end{eqnarray}
so that $D_{V_aV_b}(\pi+p)=D_{A_aA_b}(p)$. 

%%%%%%%%%%   References   %%%%%%%%%%

\end{document}